\documentstyle[fullpage,graphics,twocolumn]{article}

\title{The stabilizing role of itinerant ferromagnetism in
inter-granular cohesion in iron}

\author{D. Ye\c{s}illeten \dag,
M. Nastar \ddag, T. A. Arias \dag, A.T. Paxton \ddag, S. Yip \ddag\\
Departments of Physics\dag\  and
Nuclear Engineering\ddag\\Massachusetts Institute of Technology
Cambridge, MA 02139}

\date{\ }

\begin{document}

\maketitle

\begin{abstract}
We present a simple, general energy functional for ferromagnetic
materials based upon a local spin density extension to the Stoner
theory of itinerant ferromagnetism.  The functional reproduces well
available {\em ab initio} results and experimental interfacial
energies for grain boundaries in iron.  The model shows that
inter-granular cohesion along symmetric tilt boundaries in iron is
dependent upon strong magnetic structure at the interface, illuminates
the mechanisms underlying this structure, and provides a simple
explanation for relaxation of the atomic structure at these
boundaries.
\end{abstract}

Iron and its steel alloys, exhibit two distinct outstanding physical
properties, high strength and high magnetic response.  Little is known
about the direct relationship, if any, between the microscopic origins
of these properties.  The exchange interaction, preference for
alignment of spins to reduce inter-electronic Coulomb repulsion
through the statistical avoidance of Fermions, drives itinerant
ferromagnetism.  Fundamental topological excitations of the
crystalline lattice govern mechanical response: dislocations mediate
plasticity, grain boundaries control microstructure and inter-granular
cohesion.  In this letter we present a new model for itinerant
ferromagnetic iron and demonstrate that the exchange interaction plays
a significant stabilizing role in grain boundaries, thereby
contributing to the strength of inter-granular cohesion.

The fundamental role played by exchange in controlling the mechanical
response of magnetic materials is at present poorly understood for
lack of an appropriate microscopic theory.  Studies in iron to date
are limited to either full-blown {\em ab initio} spin-dependent
electronic structure calculations\cite{RMP} or simple interatomic
potentials such as the embedded atom method (EAM)\cite{EAM}.  While
describing all of the correct physics, {\em ab initio} calculations
treat explicitly {\em too many} degrees of freedom to allow studies of
the complex structures of all but the simplest extended defects.  For
this reason, total energy {\em ab initio} studies of iron to date have
been limited to small clusters encompassing the behavior of the
simplest ($\Sigma=3(111)$) grain
boundary\cite{KraskoGB,ExpandGB,WuFreemanOlson}.  Interatomic
potentials, while practical for complex systems, deal with {\em too
few} degrees of freedom to treat itinerant exchange properly.
Nonetheless, these potentials have been useful in studies of grain
boundaries\cite{EAMgb}, dislocations\cite{EAMdisloc} and even
fracture\cite{EAMfrac}.  Energies calculated for grain boundaries with
the EAM\cite{EAMgb}, however, are exaggerated by about a factor of two
when compared with experimental values\cite{VanVlack}, beginning
to approach even the experimental surface energy\cite{MissolEners}.
Here, we present a simple, general model which gives much better
results and reveals the mechanisms stabilizing the boundaries.

In response to the above weakness, Krasko\cite{KraskoPot} has
introduced recently an atomistic potential which includes a
prescription for estimating the exchange energy of each atom from its
local environment.  Here, we follow an alternate, microscopic route.
In accord with the microscopic origin of itinerant ferromagnetism, we
consider the extended nature of the electrons and determine only the
spin-moment coupling constant, the Stoner parameter\cite{Stoner}, as a
function of the local atomic environment.

{\em Microscopic Approach ---} Such intermediate, electronic structure
based descriptions of iron have been developed in previous studies of
iron\cite{HasePett,ZOT}, but for ideal crystalline systems.  Our
approach to the study of defects is to first identify the smallest,
physically reasonable set of degrees of freedom from these studies,
and then extend this treatment to include inhomogeneous systems.

Hasegawa and Pettifor\cite{HasePett} reproduce the experimental $P-T$
phase diagram of iron by combining a tight-binding description of the
$d-$bands with a treatment of spin-fluctuation effects.  They note
that as $T \rightarrow 0$, spin fluctuation effects become unimportant
and their spin-fluctuation theory reduces to the traditional
mean-field Stoner theory of itinerant ferromagnetism\cite{Stoner}.
They identify the temperature at which this happens to be $T_f \approx
500$~K.  Zhong, Overney, and Tomanek\cite{ZOT} take up the fact that
mean-field Stoner theory is sufficient at room temperature, and build
a model for bulk crystalline iron based upon an $spd$ tight-binding
Hamiltonian and Stoner theory, treating only the mean atomic spins.

We therefore identify a minimal set of active physical degrees of
freedom below $T_f$ to consist of the net mean spin moment on each
atom and single particle states constructed from atomic-like
$d-$orbitals.  To pass beyond perfect crystalline material, we
introduce a local atomic spin-density extension to Stoner theory.

{\em Construction of Energy Functional ---} Our selected degrees of
freedom are (1) the linear combination coefficients $\psi_{nk}(i,m)$
describing the bonding among the $m_z=-2,\ldots,2$ atomic $d-$states
of each atom $i$ for each band $n$ and each point in the Brillioun
zone $k$, (2) the net spin $n_\sigma(i)$, $\sigma=\pm 1$, associated
with each atom $i$, and (3) the location $\vec \tau_i$ of each atom
$i$.


To describe the bonding contribution to the energy $\epsilon_{nk}$
associated with each single-particle orbital, we use an orthogonal
two-center tight-binding Hamiltonian description\cite{SlaterKoster}.
In the usual fashion, we take the diagonal elements of the matrix of
hopping integrals among $L_z$ eigenstates for two atoms separated by a
distance $\tau$ along the z-axis,
$dd\delta,dd\pi,dd\sigma,dd\pi,dd\delta$, to decay exponentially with
distance, $dd\lambda=dd\lambda_o e^{-q \tau}$.  To set the primitive
matrix elements $dd\lambda_o$ and the decay factor $q$ we insist that
the tight-binding Hamiltonian reproduce well the {\em ab initio}
density of states for the $d-$bands, which is accomplished by setting
%
the Slater-Koster parameters to
$dd\sigma_o:dd\pi_o:dd\delta_o$=-1~Ryd:1~Ryd:-0.5~Ryd and the decay
factor to $q=0.63872$.
%

In summing over the single particle energies, we replace the integral
over the Brillioun zone with a discrete sum with appropriate weights
$w_k$\cite{Monk}.  The tendency of our system to spin-polarize
requires us to consider separate Fermi occupation numbers
$f_{nk\sigma}$ for each spin channel $\sigma$.  We also include the
Fermionic entropy
$$
s_{nk\sigma} \equiv -k_B \left[f_{nk\sigma} \ln f_{nk\sigma} +
(1-f_{nk\sigma}) \ln (1-f_{nk\sigma}) \right],
$$
so that our
band structure energy appears as the first sum in the energy
functional (\ref{eqn:E}) where the $\epsilon_{nk}$ are the eigenvalues
of the tight-binding Hamiltonian.
All of the results below are computed at $T=300^oK$.

Combining the Fermi occupations $f_{nk\sigma}$ with the expansion
coefficients $\psi_{nk\sigma}$ yields the local atomic spin densities,
$n_\sigma(i) \equiv \sum_{n,k,m} w_k f_{nk\sigma} \left|
\psi_{nk}(i,m) \right|^2.$ The Stoner theory of itinerant
ferromagnetism describes an energetic benefit of polarization due to
exchange of magnitude $N_{at} I m^2/4$, where $N_{at}$ is the total
number of atoms in the crystal, $m$ is the spin polarization per atom
and $I$ is the Stoner parameter\cite{Stoner}.  To extend this theory
beyond homogeneous bulk systems, we introduce a local approximation to
the exchange energy in the same spirit as the local-density
approximation\cite{KohnSham} of density functional
theory\cite{HohenbergKohn}.  In particular, for an inhomogeneous
system, we associate a separate energy contribution to the exchange
from each atom equal to what we would expect on a per atom basis from
a homogeneous system consisting of atoms in the identical environment
with identical spin polarization.  This contribution appears as the
second sum in our energy functional (\ref{eqn:E}).  The physical
motivation for this approximation is that the itinerant nature of
magnetism in iron tends to smooth variations in the spin polarization,
limiting the effects of gradient corrections.  This approach has the
advantage of allowing us to draw upon {\em ab initio} values of the
Stoner exchange parameter in bulk.

Krasko\cite{KraskoI} has performed {\em ab initio} linear response
theory calculations of the Stoner parameter $I$ in bcc and fcc bulk
iron and found it to have a mild, approximately linear volume
dependence, $dI/ds = -0.01$~Rydberg/Bohr where $s$ is the Wigner-Seitz
(WS) radius, and to have slightly different values for the bcc and fcc
lattices, $I_{bcc}^o=0.072$~Ryd/$\mu_B^2$ and
$I_{fcc}^o=0.069$~Ryd/$\mu_B^2$ at $s=2.66$~Bohr.  The use of these
values in our model gives the correct magnetic and non-magnetic states
for the bcc and fcc structures, respectively.  However, to yield the
correct total energy ordering of states, we have found necessary a
slight enhancement of the bulk Stoner parameters to
$I_{bcc}^o=0.077$~Ryd/$\mu_B^2$ and $I_{fcc}^o=0.070$~Ryd/$\mu_B^2$,
which leads to no magnetization in the fcc phase and a bcc phase
magnetic moment of $2.56 \mu_B$/atom,
somewhat enhanced relative to the accepted moment of $2.2 \mu_B$/atom.

Going beyond bulk to inhomogeneous systems with intermediate
coordinations $Q$, we make a linear interpolation for $I_i$ between
what would be expected at the same WS radius for the $Q_{bcc}=8$ and
the $Q_{fcc}=12$ lattices.  We determine the local coordination number
$Q_i$ and WS radius $s_i$ for each atom $i$ with the formulation
developed by Sawada\cite{Sawada} and the conversion $s_i =
R_i(a+b/Q_i+c/Q_i^2)/2$ from his parameter $R_i$.  We find that
setting Sawada's coefficients to $\lambda_1 \equiv
4.5023$~Bohr$^{-1}$, $\lambda_2 \equiv 10.6376$~Bohr$^{-2}$ and using
$a\equiv 1.7144$, $b\equiv -9.0948$, $c\equiv 56.372$ reproduces to
within 0.01\% the correct coordination numbers and to within 0.3\% the
correct WS radii for the diamond structure, bcc and fcc lattices
packed at the atomic density of bcc iron.  With the Stoner parameters
thus determined, we set the filling $N_d$ of the manifold of
$d-$states, so that the Fermi level for spin-down electrons in the bcc
structure falls precisely at the minimum of the pseudo-gap in the
tight-binding density of states, reproducing the physical behavior
observed in {\em ab initio} calculations.  The resulting filling,
$N_d=6.7$~electrons/atom, is in good agreement with the value of
$N_d=7.0$ used successfully in \cite{HasePett}.

Finally, for the energy associated with the locations of the atoms, we
take a power-law relationship between the interatomic potential and
the hopping elements, a standard successful practice in tight-binding
calculations\cite{PettBookHard}, to produce the final term in
(\ref{eqn:E}).  We fit the two parameters $b$ and $p$ to the
experimental equilibrium lattice constant and bulk modulus for the bcc
phase of iron, yielding $p=2.2355$~Bohr$^{-1}$ and $b=872.5174$~Ryd,
respectively.  The ratio $\lambda \equiv p/q$ corresponds to a {\em
normalized hardness} \cite{PettBookHard}, $\alpha_h \equiv
(\lambda-1)/\lambda \approx 0.7$, in line with the values near
two-thirds observed previously in tight-binding descriptions of the
transition metals \cite{PettBookTrans}.

Our final energy functional is thus,
\begin{eqnarray} 
E(\{\vec \tau_i\}) & = & {\min_{\psi,f}} ' \left\{\sum_{nk\sigma} w_k \left(
f_{nk\sigma} \epsilon_{nk}-T s_{nk\sigma}\right)\right. \nonumber \\
& &
-\frac{1}{4}\sum_i{I_i\left(n_\uparrow(i)-n_\downarrow(i)\right)^2}
\nonumber \\
&& + \left. 
\frac{b}{2} \sum_{i \ne j}{e^{-p \tau_{ij}}} \right\}, \label{eqn:E}
\end{eqnarray}
The constraints on the minimization are Fermi statistics,
$0 \le f_{nk\sigma} \le 1$
and the total number of $d-$electrons, $N_d = \sum_{nk\sigma} w_k
f_{nk\sigma}$.
This formulation is equivalent {\em in bulk systems} to the familiar
formulation of the Stoner theory in terms of a rigid shift between the
up and down electronic density of states.  The present formulation,
however, has the advantage in treating complex structures of allowing
distinct local Stoner parameters to be applied to each atom according
to its environment.  Finally, stationarity of the energy functional
with respect to the fillings
dramatically simplifies the evaluation of forces.

{\em Verification ---} To confirm the applicability of our description
to iron, we discuss briefly the comparison of our results with
available {\em ab initio} and experimental information for the bulk
crystalline phases and for grain boundaries in bcc iron.

In bulk, we reproduce the correct sequence bcc-hcp-fcc of phases
finding $E_{hcp}-E_{bcc}$=2.2~mRyd/atom and
$E_{fcc}-E_{bcc}$=6.5~mRyd/atom.  Our fcc-bcc energy difference, is on
the order of what is found in other calculations \cite{HasePett}, and
theoretical and experimental extrapolation \cite{Kauf},
\cite{Bendick}.  In agreement with {\em ab initio} calculations
\cite{Paxton,Stix}, we observe that the hcp phase is more stable than
the non-magnetic fcc phase for all values of the WS radius.  We
predict a pressure-induced phase transition from the bcc to the hcp
phase at a WS radius of $s=2.6$~Bohr, in good agreement with the {\em
ab initio} studies of \cite{Bendick,Soder}.  (Note that we did not fit
our parameters to produce the previous two properties.)  The small
energy differences among these phases opens the question of mechanical
stability.  Our Hamiltonian gives a mechanically stable ferromagnetic
bcc phase, even along the Bain transformation ($C'>0$).  Our predicted
$C'$ and $C_{44}$, which we have made no attempt to fit, are about
25\% lower than observed experimentally, corresponding to an
underestimation of about 12\% in the frequencies in the
long-wavelength portion of the phonon spectrum.

The literature presents an experimental determination of an average
typical grain boundary energy in $\alpha$-iron and {\em ab initio}
results for the spin moment distribution of the $\Sigma=5(310)$ and
$\Sigma=3(111)$ boundaries and the atomic relaxation of the
$\Sigma=3(111)$ boundary in iron.  Table \ref{tbl:trend} summarizes
our results for these and two other symmetric tilt boundaries.  Our
calculations were carried out in supercells containing two oppositely
oriented boundaries separated by at least eighteen layers of atoms.
We performed full structural and supercell relaxations of these
boundaries.

Table \ref{tbl:trend} shows that our energy results are in good
agreement with the experimental studies, which set the mean typical
grain boundary energy to be approximately
770~erg/cm$^2$\cite{VanVlack}.  The $\Sigma=3(112)$ boundary, the
coherent twin, is a known special case which is expected to be
unrepresentatively low in energy.
The magnetic and structural predictions of our energy functional are
in excellent quantitative agreement with the {\em ab initio}
predictions for the outward structural relaxation of the atomic planes
of the $\Sigma=3(111)$ boundary and for the fractional enhancement of
the spin moments of the symmetry plane for both the $\Sigma=5(310)$
and $\Sigma=3(111)$ boundaries.  (See $\Delta z$ and $\Delta m$ in
Table \ref{tbl:trend}).

\begin{figure} 
\begin{center}
\scalebox{0.40}{\includegraphics{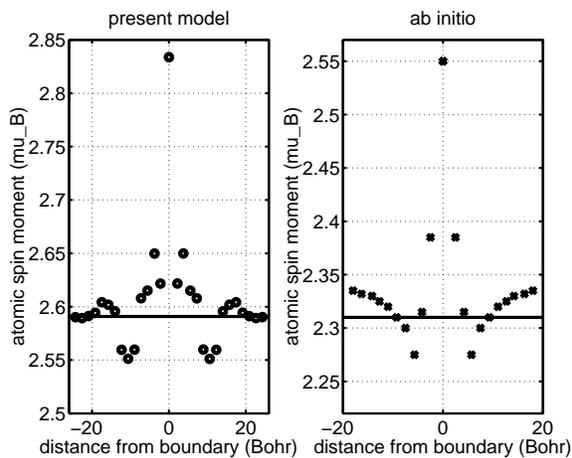}}
\end{center}
\caption{Prediction of atomic spin moments in the vicinity of the
$\Sigma=5(310)$ symmetric tilt boundary in iron.  Left: present model,
Right: {\em ab initio} results.  
Horizontal lines give the bulk phase moment of the respective calculation.}

\label{fig:moments}
\end{figure}

Finally, figure \ref{fig:moments} presents a more detailed comparison
with {\em ab initio} calculations, showing the spatial distribution of
spin moments in the vicinity of the $\Sigma=5(310)$ boundary.  The
figure shows that our model not only reproduces the enhancement of
moments on the symmetry plane but also predicts correctly the tendency
for the spin to fall below the bulk moment before eventually healing
back to the bulk value as one moves away from the boundary.  Some
discrepancies become apparent in the results at this high level level
of detail.  This comparison serves to underscores the fact that there
are limitations to any simplified, semiempirical model.  Nonetheless,
the overall level of agreement which we have found supports the
fundamental soundness of our approach and its ability to give accurate
predictions of global quantities and physical trends for complex
structures in iron.

{\em Exchange stabilization of grain boundaries in iron ---} The fact
that the embedded atom model consistently exaggerates boundary
energies in iron by a factor of two over our functional points to the
participation in boundary energetics of a microscopic mechanism more
subtle than simple distortions of the metallic bonding network.  The
data in Table \ref{tbl:trend} show that the exchange interaction is a
major factor in the physics of the grain boundaries.  The tendency to
minimize the exchange contribution drives the system to lower its
energy at the expense of an almost compensating dramatic increase in
the atomic and band structure contributions.  The end result of this
balance is the lower and much more realistic set of boundary energies
in Table \ref{tbl:trend}.  

Breakdown of the large negative contributions from the exchange
interaction shows that the enhancement comes mostly from an increase
by 10-20\% of the atomic spin moments with a lesser contribution from
an increase by 2-4\% of the Stoner exchange parameters themselves.
(Table \ref{tbl:trend}.)  The increase in magnetic moments along the
boundary therefore plays the key role in the large stabilizing
exchange interaction.

The origin of this increase is the fact that the states which
contribute to the spin moments in the system tend to be more localized
on the grain boundary.  States with energy below both Fermi levels
$\mu_\uparrow$ and $\mu_\downarrow$ are filled equally with up and
down spins and so contribute nothing to the net moment, while those
states with energy above both Fermi levels are completely empty.  Only
those states between these two Fermi levels are filled with
uncompensated spins and contribute to the spin moments.  The range of
energies between the two Fermi levels is near the center of the
$d-$band.  The states in this energy range tend to localize on the
boundary as a direct consequence of the more open structure of the
boundary, which lowers the tight-binding matrix elements $dd\lambda$
and thereby narrows the band toward the band-center.   We confirmed
this latter behavior by direct inspection of the electronic states.

Finally, our model also sheds light on the geometric relaxation of the
grain boundaries.  In all boundaries in our study, the two planes of
atoms immediately neighboring the symmetry plane to relax outward and
compress into the surrounding bulk.  Because the Stoner parameters
increase with decreasing atomic volume ($dI/ds<0$, above) this
diminishes the Stoner parameter on the symmetry plane but enhances the
exchange parameter in a total of four planes, one pair on either side
of the boundary.  This relaxation pattern enhances the exchange
stabilization by sacrificing the exchange parameters on the single
symmetry plane in favor of a total of four nearby planes.

In conclusion, we have developed a local spin density functional
description of itinerant ferromagnetic materials which provides a
simple and accurate picture of the relationships among geometry,
electronic structure and stability of tilt grain boundaries in iron.

\begin{center}{\bf Acknowledgments}\end{center}
This work was supported by the MRSEC Program of the National
Science Foundation (DMR 94-00334) and by the Alfred
P. Sloan Foundation (BR-3456).  Computational support
provided by the MIT Xolas prototype SUN cluster.

\newpage

\begin{table}
\begin{center}
\begin{tabular}{cccccc}
Boundary & $E_{gb}$ & $E_{xc}$ & $\Delta m$ & $\Delta I_{max}$ & $\Delta z$ \\
     & erg/cm$^2$ & erg/cm$^2$ &      &            &  Bohr \\  \hline
$\Sigma=3(112)$ & 140 & -8600 & 14\% & 2\% & 0.1 \\
          & (300$^{**}$) & & & & \\ \hline
$\Sigma=5(310)$ & 560 & -3800 &  9\%  & 3\% & 0.5 \\
                & (1300$^{**}$) &       & (8\%$^\dag$) & & \\ 
                & ($\sim 770^{*}$) & & & & \\ \hline
$\Sigma=3(111)$ & 770                   & -5200 & 18\% & 4\% & 0.6 \\
& & & (15$^\ddag$-18\%$^+$) & & (0.5-0.8$^+$) \\ \hline
$\Sigma=9(114)$ & 760 & -4100 & 15\% & 4\% & 0.9 \\
          & (1450$^{**}$) & & & & \\ 
                & ($\sim 770^{*}$) & & & & \\ \hline \hline
\multicolumn{6}{l}{Expt: $^{*}$\cite{VanVlack}, EAM: $^{**}$\cite{EAMgb}; {\em Ab initio}:
     $^\dag$\cite{Sigma5}, $^\ddag$\cite{KraskoGB}, $^+$\cite{WuFreemanOlson}.}
\end{tabular}
\end{center}
\caption{Summary of Grain Boundary Results: Boundary formation energy
[$E_{gb}$], exchange contribution to the energy [$E_{ex}$], change in
magnetic moment on the boundary plane [$\Delta m$], maximum Stoner
parameter [$\Delta I_{max}$], outward motion of planes immediately
neighboring the boundary [$\Delta z$].  Results of other studies
appear in parenthesis.}
\label{tbl:trend}
\end{table}

\end{document}